\documentclass[sigconf,screen, nonacm]{acmart}
\usepackage{graphicx}
\usepackage{xcolor}
\usepackage{todonotes}

\usepackage{enumitem}
\usepackage{url}
\usepackage{balance}
\usepackage{flushend}
\begin{document}

\title{SQuaD: The Software Quality Dataset}
\subtitle{A Multi-Dimensional Time-Aware Collection of Metrics for Large-Scale Empirical Research}



\author{Mikel Robredo}
\affiliation{%
  \institution{University of Oulu}
  \city{Oulu}
  \country{Finland}
}
\email{mikel.robredomanero@oulu.fi}

\author{Matteo Esposito}
\affiliation{%
  \institution{University of Oulu}
  \city{Oulu}
  \country{Finland}
}
\email{matteo.esposito@oulu.fi}

\author{Davide Taibi}
\affiliation{%
  \institution{University of Southern Denmark }
  \city{Vejle}
  \country{Denmark}
}
\additionalaffiliation{%
  \institution{University of Oulu, Finland}
  \city{Oulu}
  \country{Finland}
}
\email{taibi@imada.sdu.dk}

\author{Rafael Peñaloza}
\affiliation{%
  \institution{University of Milano-Bicocca}
  \city{Milan}
  \country{Italy}
}
\email{rafael.penalozanyssen@unimib.it}

\author{Valentina Lenarduzzi*}
\affiliation{%
  \institution{University of Southern Denmark}
  \city{Vejle}
  \country{Denmark}
}

\email{lenarduzzi@imada.sdu.dk}

\begin{abstract}
\sloppy Software quality research increasingly relies on large-scale datasets that measure both the product and process aspects of software systems. However, existing resources often focus on limited dimensions, such as code smells, technical debt, or refactoring activity, thereby restricting comprehensive analyses across time and quality dimensions. To address this gap, we present the Software Quality Dataset (SQuaD), a multi-dimensional, time-aware collection of software quality metrics extracted from 450 mature open-source projects across diverse ecosystems, including Apache, Mozilla, FFmpeg, and the Linux kernel. By integrating nine state-of-the-art static analysis tools, i.e., SonarQube, CodeScene, PMD, Understand, CK, JaSoMe, RefactoringMiner, RefactoringMiner++, and PyRef, our dataset unifies over 700 unique metrics at method, class, file, and project levels. Covering a total of 63,586 analyzed project releases, SQuaD also provides version control and issue-tracking histories, software vulnerability data (CVE/CWE), and process metrics proven to enhance Just-In-Time (JIT) defect prediction. The SQuaD enables empirical research on maintainability, technical debt, software evolution, and quality assessment at unprecedented scale. We also outline emerging research directions, including automated dataset updates and cross-project quality modeling to support the continuous evolution of software analytics. The dataset is publicly available on ZENODO (DOI: 10.5281/zenodo.17566690).


\end{abstract}



\maketitle

\section{Introduction}
\label{sec:intro}

Software maintenance is a core facet of Software Quality (SQua) as it helps teams to extend and correct their software system more easily~\citep{fowler1999refactoring}. The Software Engineering (SE) community considers multiple factors and mechanisms that affect and help to improve the software quality of a system~\citep{fowler1999refactoring, lenarduzzi2018survey}. These facets can range from software vulnerabilities~\citep{esposito2024extensive} to code quality issues~\citep{esposito2025correlation}, as well as technical debt management~\citep{lenarduzzi2019diffuseness} and code refactoring operations~\citep{silva2016we, robredo2024analyzing}, among others. 

Empirical mining software repository research relies on the public availability of open source software repositories hosted in platforms such as GitHub, and on the correctness of the mining activity is performed~\citep{kalliamvakou2014promises}. Based on this premise, researchers often make significant efforts to select subsets of projects based on code quality~\citep{esposito2024extensive, sharma2021qscored}, as well as on the self-implemented \textit{codes of conduct} in software foundations such as the \textit{Apache Software Foundation} (ASF)~\citep{lenarduzzi2019technical}.\footnote{\url{https://www.apache.org/foundation/}} Similarly, a popular technique for ensuring the quality of a software system during its development and maintenance is the employment of Static Analysis Tools (SAT)~\citep{esposito2025correlation}. Multiple studies have exploited the use of SATs to remediate common quality issues~\citep{esposito2024extensive, palomba2016smells, taibi2017developers, lenarduzzi2020some}. Most of the times, researchers concentrate on a specific set of projects and SATs due to the resource-intensive and time-consuming task of employing a larger number of SATs on a large-scale set of projects. 

Consequently, existing works already provide the research community with large-scale datasets to enable researchers to answer potential research questions by leveraging the shared data. For instance, \citet{lenarduzzi2019technical} contributed to the SE literature with a large-scale dataset on Technical Debt (TD) metrics derived from SATs like SonarQube (SQ), later expanded by \citet{graf2024different}. In addition, further research efforts have been made to contribute to the SE community with datasets on software quality aspects such as software vulnerabilities~\citep{bui2022vul4j}, code smells and quality metrics~\citep{sharma2021qscored}, time series-based software evolution metrics~\citep{sousa2022time}, as well as code refactoring activity~\citep{kadar2016code}, among others~\citep{kishimoto2025dataset, hourri2025dataset}. 

However, to date, no existing works combine all these SQua aspects into a single large-scale dataset. To such end, we leveraged nine state-of-the-art SATs to mine mature SE projects from sources such as the ASF, the \textit{Mozilla}~\footnote{\url{https://www.mozillafoundation.org/en/}}, the \textit{FFMpeg}~\footnote{\url{https://ffmpeg.org}} foundations and the \textit{Linux kernel}~\footnote{\url{https://www.kernel.org/doc/html/latest/}}. We employed \textbf{SQ} and \textbf{CodeScene} (CS) to evaluate TD and security issues~\citep{lenarduzzi2019technical, tornhill2018assessing}, offering insights into maintainability and code health. Covering the aspect of code refactoring, we adopted \textbf{RefactoringMiner} (RMiner)~\citep{tsantalis2020refactoringminer}, \textbf{RefactoringMiner++} (RMiner++)~\citep{ritz2025refactoring} and \textbf{PyRef}~\citep{atwi2021pyref} for Java, C++ and Python languages accordingly. We assessed the \textit{coding rule compliance} using \textbf{PMD}~\citep{esposito2025correlation, esposito2024extensive} and \textbf{Understand}~\citep{bakhtin2025network} by SciTools. We mined product and process metrics at different granularity levels with \textbf{CK}~\citep{aniche-ck} and \textbf{JaSoMe}~\citep{jasome-tool}. We expanded the mined data with up-to-date issue reports from GitHub, Jira and BugZilla issue trackers (ITS), their reported Common Vulnerabilities and Exposures (CVE), and Common Weakness Enumeration (CWE) types with their official definitions as well as additional process metrics demonstrated to improve JIT prediction accuracy~\citep{falessi2021impact, madeyski2015process, falessi2023enhancing}.

\sloppy Thus, in this paper, we present \textit{the Software Quality dataset} (SQuaD), which provides the community with a multi-dimensional time-aware collection of metrics for large-scale empirical research.
The main contributions of this paper are:

\begin{itemize}
    \item \textit{The SQuaD}. A large-scale set of 450 projects where, by leveraging nine state-of-the-art SATs, we analyzed 725 metrics describing common SQua aspects from all the versions of their officially reported releases, covering metrics at method, class, file and project level.
    \item \textit{Two data formats}. CSV files and a noSQL database, thus enabling researchers to access our dataset efficiently.
    \item The replication package with the scripts to use the SATs that produced this dataset.
\end{itemize}

\textbf{Paper Structure.} Section~\ref{sec:construction} describes the construction method adopted for this dataset. Section~\ref{sec:structure} presents the dataset and its usage. Section~\ref{sec:impact} highlights the future research opportunities using the dataset can provide. Section~\ref{sec:limitations} acknowledges the limitations of the dataset. Section~\ref{sec:conclusion} draws conclusions and future works. 


\section{Dataset construction}
\label{sec:construction}

\begin{figure*}[t]
    \centering
    \includegraphics[width=0.85\textwidth]{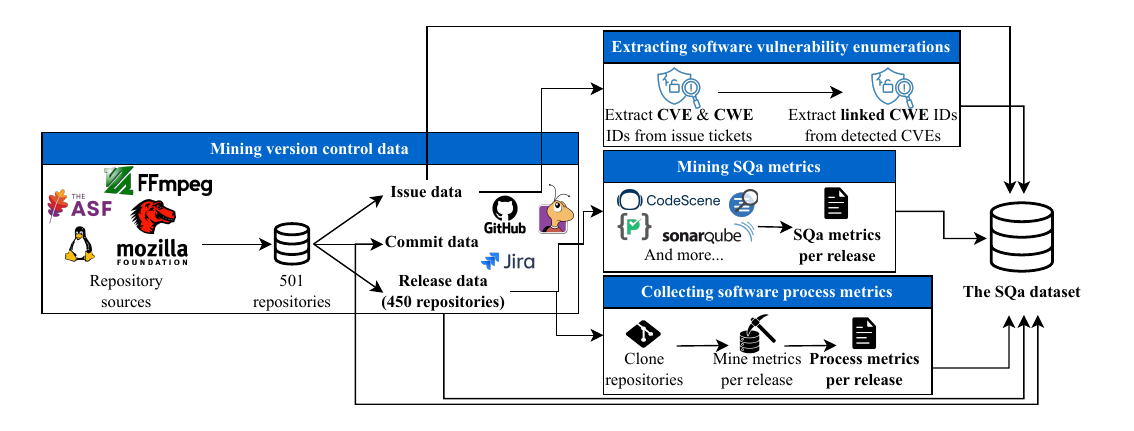}
    \caption{Overview of the dataset construction methodology.}
    \label{fig:mining-arch}
\end{figure*}

This section describes the data sources used to create the dataset, and the methodology used to gather the data, which we graphically present in Figure~\ref{fig:mining-arch}. The construction of the dataset required four main data mining stages: \textit{Mining version control data}, \textit{Mining SQua metrics from the selected SATs}, \textit{extracting software vulnerability enumerations}, and \textit{collecting software process metrics}.

\subsection{Mining version control data}
\label{sec:mining-version-control}

To collect the initial set of software repositories to include in our dataset, we considered mining classically investigated projects from sources such as the ASF, the Mozilla Foundation, and the Linux kernel~\citep{lenarduzzi2019technical}. We applied an additional filtering process to include only active, mature projects~\cite{kalliamvakou2014promises, saarimaki2025does, amoroso2024dataset}. For that, we excluded \textbf{archived} projects or \textbf{based on forks}, as well as projects with \textbf{no available SBOM}. Furthermore, we excluded projects with \textbf{no activity in the last six months}, projects that had \textbf{less than three contributors}, and those that had \textbf{less than 50 stars on GitHub}.

We leveraged GitHub's API\footnote{\url{https://docs.github.com/en/rest?apiVersion=2022-11-28}} to mine their commit history as well as the issue tracking history for those projects that used GitHub as their ITS. We also mined the issue tracking history from those ASF reporting to use Jira\footnote{\url{https://developer.atlassian.com/cloud/jira/platform/rest/v3/intro/\#about}} and BugZilla\footnote{\url{https://bugzilla.readthedocs.io/en/5.2/api/}} as their official ITS\footnote{\url{https://issues.apache.org}}. With a total of 501 software repositories detected in the selected sources of data, only 450 reported published releases or tags in GitHub, which we set as the observational points to mine the SQua metrics, and thus build the historical development progress from the mined repositories.



\subsection{Mining SQua metrics}
\label{sec:mining-sqa}

This section describes the systematic methodology used to mine software metrics from the adopted SATs. Since each tool captures distinct aspects of software quality, Table~\ref{tab:sqa-tools} reports the number of metrics extracted per SAT and their covered dimensions. We provide instructions on replicating our mining pipeline in the replication package~\cite{robredo2025squad-replication}.

\begin{table}[htb]
\centering
\small
\caption{Overview of adopted SATs, mined metrics, and aspects analyzed by each SAT.}
\begin{tabular}{p{1.3cm}p{1cm}p{5cm}}
\hline 
\textbf{Tool} & \textbf{\#Metrics} & \textbf{Aspect covered (Metrics reference)} \\ 
\hline 
CK & 88 & Calculates class-level and method-level code metrics in Java projects.~\cite{aniche-ck} \\ 
JaSoMe & 70 & Mines file, package, class \& method quality metrics in Java projects.~\cite{jasome-tool} \\ 
RMiner & 103 & Detects refactorings applied in the history of a Java project.~\cite{tsantalis2020refactoringminer}\\ 
RMiner++ & 16 & Detects refactorings applied in the history of a C++ project.~\cite{ritz2025refactoring} \\ 
Understand & 111 & Mines file, class \& entity quality metrics for multiple languages.~\cite{rahman2013and} \\ 
SQ & 192 & Calculates several quality metrics \& verifies the code’s compliance against a specific set of ``coding rules''.~\cite{lenarduzzi2019technical} \\ 
PMD & 114 & Runs coding rules against source files to find violations.~\cite{esposito2025correlation} \\ 
PyRef & 9 & Detects refactorings applied in the history of a Python project.~\cite{atwi2021pyref} \\ 
CodeScene & 22 & Computes per-file comprehensive code health checks.~\cite{tornhill2018assessing} \\ 
\hline 
\end{tabular}
\label{tab:sqa-tools}

\end{table}

Since each SAT required a different mining setting, we followed a systematic mining approach for each SAT in parallel. 1) We cloned the software repository, and subsequently, 2) we looped through the project's release commit hashes and checked out the cloned repository to the release version accordingly. 3) For each release iteration, we launched the SAT and mined the entire codebase of the repository. 

With all the repositories mined for a specific SAT, we merged the outcome from all projects into a single CSV table. Since each of the adopted SATs mined the repositories at different granularity levels, we specify the granularity level, i.e., the analyzed object type per row, within the shared replication package.

In addition, and since SQ also reports coding issues based on the codebase's compliance against SQ's "coding rules"~\cite{lenarduzzi2020some}, we leveraged SQ's API to retrieve all the raised issues for each of the release versions accordingly.\footnote{\url{https://docs.sonarsource.com/sonarqube-server/extension-guide/web-api}}

\subsection{Extracting software vulnerability enumerations}
\label{sec:extractining-cve-cwe}
From the mined issue tracking data, we searched for all the regular expressions matching the pattern for official Software Vulnerability Enumerations (CVE) and that of software weaknesses, i.e., \verb|CVE-\d{4}-\d{4,7}| and \verb|CWE-\d{3,4}|. Subsequently, we collected all the publicly available information from the official Common Weakness Enumeration (CWE) index\footnote{\url{https://cwe.mitre.org/index.html}} and the National Institute of Standards and Technology (NIST)\footnote{\url{https://nvd.nist.gov/vuln}} for each of the matched enumerations. For that, on the one hand, we leveraged the available datasets with the official information for the currently existing CWEs and fetched the information about the matched CWEs. Similarly, we used the API access provided by the NIST and retrieved the information regarding each of the matched CVEs.

\subsection{Collecting software process metrics}
\label{sec:collecting-sw-process}

Recent research efforts have demonstrated that specific process metrics are more helpful than the structure of the source code itself when training JIT defect prediction models~\citep{falessi2021impact, madeyski2015process, falessi2023enhancing}. Since we already collected all the characteristics representing the structure of the code base throughout the entire release history of the mined repositories, we are now interested in collecting software process metrics at each release version of the projects. For that, we used Python's \texttt{GitPython}\footnote{\url{https://gitpython.readthedocs.io/en/stable/}} library to safely traverse through the entire version control history of the cloned repositories. Thus, we collected the process metrics highlighted in the literature at each release version.





\section{The Software Quality dataset}
\label{sec:structure}
The SQuaD comprises measures for a total number of 725 SQua metrics distributed across the employed 9 state-of-the-art SATs. These results represent metric observations from a total of 63,586 analyzed project releases and tags, based on a total number of 450 software repositories. The dataset contains a total of 628,178 defect tickets, 2,622,413 GitHub commits, and official information on 1479 CVE and 175 CWE enumerations detected within the mined issue tickets. Furthermore, the dataset provides the computed value of 14 process metrics covering the entire version control history of the mined projects. On average, the projects included are over 9 years old, with a mean number of 125,500 total lines of code, 2465 stars in GitHub, and over 104 contributors per project.


The dataset is stored in two different formats. We utilize a NoSQL database, specifically MongoDB.\footnote{\url{https://www.mongodb.com}} We facilitate access to this format of the database through the \textit{Binary JSON} (BSON) format \footnote{\url{https://www.mongodb.com/resources/languages/bson}}, the standard sharing format in MongoDB, and compressed via \textit{Z-standard} \cite{collet2018zstandard}. Similarly, we provide the dataset in a series of CSV files following the same entity relationship designed for the database format. We include an entity diagram with the table hierarchy in the shared replication package~\cite{robredo2025squad-replication} to facilitate its use.

\begin{itemize}
    \item Table \texttt{projects\_data} contains the links to the GitHub repository.
    \item Table \texttt{COMMITS} reports the commit information retrieved from GitHub, including the commit hash, the commit message, the commit date and the alias of the commit author, among other attributes.
    \item Table \texttt{ISSUES} contains the issue tickets from the mined projects. Based on the column \texttt{its}, the table provides details about issues registered in GitHub, Jira and Bugzilla. 
    \item Table \texttt{release\_data} contains the identifier of the project releases and tags retrieved from GitHub as well as their related commit hash.
    \item Table \texttt{summary\_statistics} contains summary statistics retrieved from GitHub, such as the number of stars, the number of contributors, or the number of watchers, among others.
    \item Table \texttt{PRJ\_ITS\_VLN\_LINKAGE} contains the linkage between project identifiers, issue trackers, and detected vulnerability references.
    \item Table \texttt{cwe\_data} contains the official information retrieved from the CWE official index on the enumerations detected in the issue tickets of the mined projects.
    \item Table \texttt{cve\_data} contains the official information retrieved from the NIST on the enumerations detected in the issue tickets of the mined projects. Moreover, it also contains information on the enumerations related to the CWE weaknesses detected.
    \item Table \texttt{process\_metrics} provides the computed values of the collected process metrics for all the releases mined from the projects included in the dataset.
    \item The \texttt{TOOL} tables consist a table per each SAT used during the mining process. Each of the tables contains observations at different granularity levels, in some cases uniform during the entire table (e.g. refactoring observations for RMiner), and in other cases at different granularity levels (e.g. file, class and method level for JaSoMe), always specifying the metric type through the \texttt{metric} column.
\end{itemize}

We made the dataset as well as the raw mined data  accessible~\cite{robredo2025squad-zenodo} in \texttt{ZENODO}. The compressed BSON database can be imported into MongoDB and thus be explored through the MongoDB shell or any other graphical interface supporting MongoDB. We also provide the dataset in CSV format, thus facilitating one CSV file per table listed above.


\section{Impact and potential research directions}
\label{sec:impact}

Software quality metrics stand as one of the most important source of information that can describe the development process of a project~\cite{lenarduzzi2019technical}. Consequently, this dataset stands as the largest dataset release till the date, combining SQua metrics from some of the state-of-the-art SATs employed for measuring SQua.

The SQuaD opens a wide availability of data for multiple potential use cases. Researchers can investigate time-dependent trends and variables over different granularity levels~\cite{robredo2024analyzing}, for instance, in order to perform software evolution and change analysis. Similarly, multiple studies could leverage our dataset to benchmark different technical debt indicators such as code smells across different ecosystems (e.g. ASF, Linux Kernel).

Since the SQuaD integrates CVE/CWE and issue-tracking data, researchers investigating defect prediction can use our dataset to test further novel prediction models~\cite{esposito2024extensive, madeyski2015process}, as well as forecasting models that might require the data to be already chronologically ordered. Building upon this, with the surge of models enabled by the \textit{Transformer} architecture~\cite{han2025transformer, vaswani2017attention}, prediction models require a larger dimension of data for training. The SQuaD stands as a potential candidate to provide this capability to SE researchers.

\section{Limitations}
\label{sec:limitations}

The creation of the SQuaD involved using some of the some of the most commonly used SATs. We are aware that these tools might analyze the code incorrectly under some conditions, especially when the programming language is structurally different. Hence, we aimed at only adopting state-of-the-art SATs to reduce this limitation. Similarly, the tools PyRef and RMiner++ generated multiple compatibility issues when including them in the mining pipeline, hence the smaller size of their mined results. We relate this limitation to their novelty of their release, aiming to export the model of RefactoringMiner to other programming language.

Another important, yet controversial limitation of the dataset is its size. We aimed at mining some of the open-source repositories closely related to industry projects, and therefore, the dimension of the mined output resulted in a dimension that will require practitioners to have powerful machines to enable the use of the SQuaD.

\section{Conclusion}
\label{sec:conclusion}

In this work, we presented the SQuaD dataset. It stands as the largest source code dataset analyzing software projects based on different programming languages, and mined with SATs widely used in industry and research.

We described the dataset construction process to mine the data. We provided the SQuaD in CSV and BSON compressed formats to facilitate the compact use of the data. The SQuaD includes mined results of 725 SQua metrics from 9 different SATs, collected from the project versions across 450 software projects. The creation of the SQuaD required 7 months of mining process due to the license limitations of some other tools, as well as due to the size of some of the mined projects. The provided data allows researchers to perform large-scale studies without dealing with the data collection process, but directly fetch the data they need from the SQuaD and conduct the study.

Our plans involve expanding and updating the SQuaD by including new repositories, releasing the database in MySQL format, and expanding the tool selection.

\section*{Acknowledgment}

The authors wish to acknowledge \textbf{CSC—IT Center for Science}, Finland, for generous computational resources, specifically the Mahti supercomputer, the Allas cloud storage system, and the cPouta cloud community service. Similarly, this work has been funded by \textbf{FAST}, the Finnish Software Engineering Doctoral Research Network,
funded by the Ministry of Education and Culture, Finland. Lastly, the authors wish to acknowledge \textbf{SciTools} for their constant availability and assistance for using their software \textbf{Understand}.

\balance
\bibliographystyle{ACM-Reference-Format}
\bibliography{main}

\end{document}